\documentclass[12pt]{iopart}

\usepackage{graphicx}
\usepackage{color}

\begin{document}

\title[L\'evy processes on a generalized fractal comb]{L\'evy processes on a generalized fractal comb}

\author{Trifce Sandev$^{\dagger,\ddagger}$, Alexander Iomin$^{\P}$, and Vicen{\c c} M\'{e}ndez$^{\flat}$} 
\address{$\dagger$ Max Planck Institute for the Physics of Complex Systems,
N\"{o}thnitzer Strasse 38, D-01187 Dresden, Germany\\
$\ddagger$ Radiation Safety Directorate, Partizanski odredi 143, P.O. Box 22,
1020 Skopje, Macedonia\\
$\P$ Department of Physics, Technion, Haifa 32000, Israel\\
$\flat$ Grup de F\'{\i}sica Estad\'{\i}stica, Departament
de F\'{\i}sica. Universitat Aut\`onoma de Barcelona. Edifici Cc.
08193 Cerdanyola (Bellaterra) Spain}


\begin{abstract}
Comb geometry, constituted of a backbone and fingers, is one of the most simple paradigm of a two dimensional structure, where anomalous diffusion can be realized in the framework of Markov processes. However, the intrinsic properties of the structure can destroy this Markovian transport. These effects can be described by the memory and spatial kernels. In particular, the fractal structure of the fingers, which is controlled by the spatial kernel in both the real and the Fourier spaces, leads to the L\'evy processes (L\'evy flights) and superdiffusion. This generalization of the fractional diffusion is described by the Riesz space fractional derivative. In the framework of this generalized fractal comb model, L\'evy processes are considered, and exact solutions for the probability distribution functions are obtained in terms of the Fox $H$-function for a variety of the memory kernels, and the rate of the superdiffusive spreading is studied by calculating the fractional moments. For a special form of the memory kernels, we also observed a competition between long rests and long jumps. Finally, we considered the fractional structure of the fingers controlled by a Weierstrass function, which leads to the power-law kernel in the Fourier space. It is a special case, when the second moment exists for superdiffusion in this competition between long rests and long jumps.
\end{abstract}


\submitted{J. Phys. A: Math. Theor.}

\section{Introduction}

A comb model is a particular example of a non-Markovian motion, which takes place due to its specific geometry realization inside a two dimensional structure.
It consists of a backbone along the structure $x$ axis and fingers along the $y$ direction, continuously spaced along the $x$ coordinate, shown in Fig. 1. This special geometry has been introduced to investigate anomalous diffusion in low-dimensional percolation clusters \cite{arkhincheev,havlin,weiss,white}. In the last decade the comb model has been extensively studied to understand different realizations of non-Markovian random walks, both continuous \cite{arkhincheev1,baskin iomin prl,lenzi} and discrete \cite{cassi}. In particular, comb-like models have been used to describe turbulent hyper-diffusion due subdiffusive traps \cite{baskin iomin prl,iomin baskin}, anomalous diffusion in spiny dendrites \cite{iomin,mendez}, subdiffusion on a fractal comb \cite{iomin2}, and diffusion of light in L\'evy glasses \cite{nature06948} as L\'evy walks in quenched disordered media \cite{BurPRE81,BurPRE89}, and to model anomalous transport in low-dimensional composites \cite{baklanov}. 

\begin{figure}\label{fig1}
\centering{\includegraphics[width=8cm]{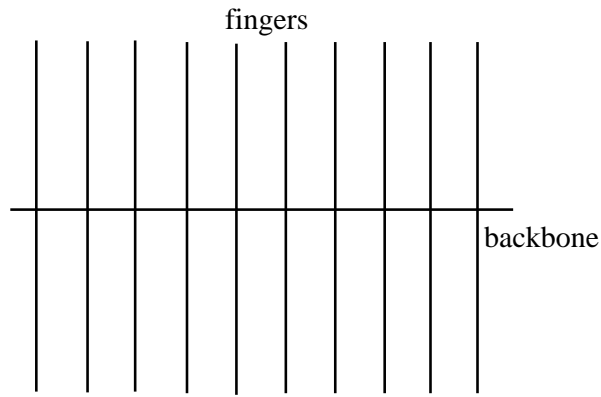}
\caption {Comb-like structure.}}
\end{figure}

The macroscopic model for the transport along a comb structure is presented by the following two-dimensional heterogeneous diffusion equation \cite{arkhincheev,havlin,weiss,white}
\begin{eqnarray}\label{classical comb}
\frac{\partial}{\partial t}P(x,y,t)
=\mathcal{D}_{x}\delta(y)\frac{\partial^{2}}{\partial x^{2}}P(x,y,t)+\mathcal{D}_{y}\frac{\partial^{2}}{\partial y^{2}}P(x,y,t),
\end{eqnarray}
where $P(x,y,t)$ is the probability distribution function (PDF),
$\mathcal{D}_{x}\delta(y)$ and $\mathcal{D}_{y}$ are diffusion coefficients in the $x$ and $y$ directions, respectively, with physical dimension
$[\mathcal{D}_{x}]=\mathrm{m}^{3}/\mathrm{s}$, and $[\mathcal{D}_{y}]=\mathrm{m}^{2}/\mathrm{s}$. The $\delta(y)$ function (the Dirac $\delta(y)$ function) means that diffusion in the $x$ direction occurs only at $y=0$. This form of equations describes diffusion in the backbone (at $y=0$), while the fingers play the role of traps. Diffusion in a continuous comb can be described within the continuous time random walk (CTRW) theory \cite{report}. For the continuous comb with infinite fingers, the returning probability scales like $t^{-1/2}$, and the waiting time PDF behaves as $t^{-3/2}$ \cite{metzler cppc}, resulting in appearance of anomalous subdiffusion along the backbone with the transport exponent $1/2$. In another example of a fractal volume of an infinite number of backbones, it has been shown that the transport exponent depends on the fractal dimension of the backbone structure \cite{sandev iomin kantz}. Natural phenomenological generalization of the comb model (\ref{classical comb}) is the generalization of both the time processes, by introducing memory kernels $\gamma(t)$ and $\eta(t)$, and introducing space inhomogeneous (fractal) geometry, i.e., a power-law density of fingers described by kernel $\rho(x)$ \cite{iomin2,iomin3,iomin}. This modification of the comb model (\ref{classical comb}) can be expressed in the form of a so-called fractal comb model
\begin{eqnarray}\label{diffusion like eq on a comb0}
\int_{0}^{t}dt'\,\gamma(t-t')\frac{\partial}{\partial t'}P(x,y,t')
&=\mathcal{D}_{x}\delta(y)\int_{0}^{t}dt'\,\eta(t-t')\frac{\partial^{2}}{\partial x^{2}}P(x,y,t')\nonumber\\&+\mathcal{D}_{y}\frac{\partial^{2}}{\partial y^{2}}\int_{-\infty}^{\infty}dx'\,\rho(x-x')P(x',y,t).
\end{eqnarray}
Here, the memory kernels $\gamma(t)$ and $\eta(t)$ are, in general case, decaying functions, approaching to zero in the long time limit (see \cite{joint MMNP} for details of the form of the memory kernels). The physical dimensions of the diffusion coefficients $\mathcal{D}_{x}\delta(y)$ and $\mathcal{D}_{y}$ depend now on the form of the memory kernels $\gamma(t)$ and $\eta(t)$. The memory kernels $\gamma(t)$ and $\eta(t)$, and the kernel $\rho(x)$\footnote{Note that the density of fingers is $\int dx\,\rho(x)$.} should be introduced in such a way that these functions do not change the physical meaning of the diffusion coefficients $\mathcal{D}_x\delta(y)$ and $\mathcal{D}_y$. Therefore, it is reasonable to  introduce these functions
in the dimensionless form, by introducing the time scale $\tau$ and the coordinate scale $l$. For example, it can be done in the following way \cite{iomin baskin}: $\tau=\mathcal{D}_x^2/\mathcal{D}_y^3$ and $l=\mathcal{D}_x/\mathcal{D}_y$, where we use that the dimension of $\mathcal{D}_x$ is $[\mathcal{D}_x]=l^3/\tau$, while the dimension of $\mathcal{D}_y$ is $[\mathcal{D}_y]=l^2/\tau$. This yields the corresponding change of the kernels $\gamma(t/\tau)$, $\eta(t/\tau)$, and $\rho(x/l)$, and this leads to the rescaling of Eq.~(\ref{diffusion like eq on a comb0}). To avoid this procedure and keep the diffusion parameters $\mathcal{D}_x$ and $\mathcal{D}_y$ explicitly, we just state that the diffusion coefficients automatically absorb these scale parameters, and this rescaling depends on the functional form of $\gamma(t)$, $\eta(t)$ and $\rho(x)$. The function $\gamma(t)$ contributes to the memory effects in such a way that the particles moving along the $y$-direction, i.e., along the fingers, can be trapped. It means that diffusion along the $y$ direction can be anomalous as well \cite{mendez,sandev draft}. The function $\eta(t)$ is a so-called generalized compensation kernel \cite{mendez}. The case $\gamma(t)=\eta(t)=\delta(t)$ yields the diffusion equation of the comb model (\ref{classical comb}). Corresponding CTRW models have been suggested, where the memory kernels appear in the waiting time  \cite{mendez,joint MMNP,sandev draft}. A mesoscopic  mechanism of this CTRW phenomenon has been suggested in \cite{mich2015}, as well.

The spatial fractal geometry is taken into consideration by the fractal dimension of the finger volume (mass) $|x|^{\nu}$, where $0<\nu<1$ is the fractional dimension, and fingers are continuously distributed by the power-law. This can be presented as a convolution integral between the non-local density of fingers and the PDF $P(x,y,t)$ in the form \cite{iomin2} $\int_{-\infty}^{\infty}dx'\,\rho(x-x')P(x',y,t)$, which also can be presented by the inverse Fourier transform
$\mathcal{F}_{\kappa_x}^{-1}\left[|\kappa_x|^{1-\nu}\hat{P}(\kappa_x,y,t)\right]$, where $\mathcal{F}_{x}\left[\rho(x)\right]=\tilde{\rho}\left(\kappa_{x}\right)=|\kappa_{x}|^{1-\nu}$\footnote{The Fourier transform of $f(x)$ is given by $\tilde{F}(\kappa)=\mathcal{F}
\left[f(x)\right]=\int_{-\infty}^{\infty}{d}x\,f(x)e^{\imath\kappa x}$.
Consequently, the inverse Fourier transform is defined by
$f(x)=\mathcal{F}^{-1}\left[\tilde{F}(\kappa)\right]=\frac{1}{2\pi}\int_{-\infty}
^{\infty}{d}\kappa\,\tilde{F}(\kappa)e^{-\imath\kappa x}$.}. This integration also establishes a link between fractal geometry and fractional integro-differentiation \cite{LM,nigmatulin,rutman} (see also the discussion in Summary).

As an illustration, a fractal comb is given in Fig. 2. The fractal comb in Fig. 2 is a random form of a middle third Cantor set construction, where a given segment with fingers is randomly divided in three parts and we delete the middle part. Therefore, we obtain the first generation which consists of two subsets of fingers. We repeat this middle third  procedure for each subset to obtain the second generation with four random subsets of continuously distributed fingers. Then, one obtains the third generation, an so on. One should recognize that a random walk on this fractal comb (either random or regular) leads to correlations, related to quenched structures \cite{report}. Therefore, the random structure of the comb induces correlation between successive trapping times in the fingers. In some cases of large scales, such random walks, can be renormalized to a CTRW model, and the quenched aspect can be neglected by using an effective trapping time PDF, as discussed in Ref. \cite{report}.

\begin{figure}\label{figure2}
\centering{\includegraphics[width=10cm]{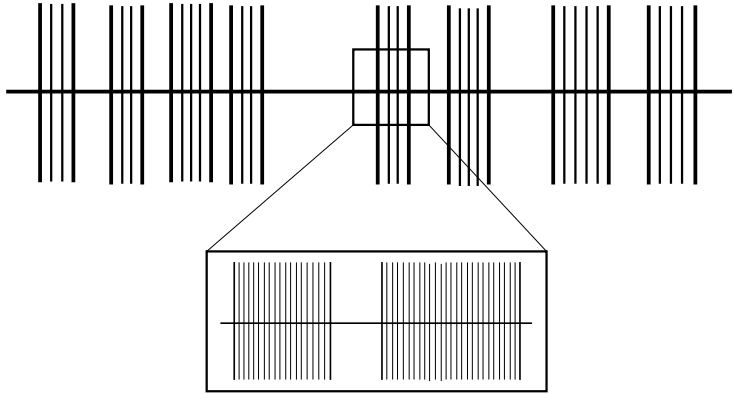}
\caption {Fourth generation of a random one-third Cantor set.
This fractal comb is a form of a middle third Cantor set construction \cite{falconer}, where each segment is randomly divided in three parts. The second slice is the first generation of the smallest part of the third generation of the Cantor set, shown in the upper slice.}}
\end{figure}
Comb model (\ref{diffusion like eq on a comb0}) for $\nu=1$ reduces to a generalization of continuous comb model for anomalous and ultraslow diffusion. Furthermore, for $\gamma(t)=\eta(t)=\delta(t)$ the ``classical'' comb model (\ref{classical comb}) is recovered, as well. The anomalous diffusion processes are characterized by power-law dependence of the mean square displacement (MSD) on time $\left\langle x^{2}(t)\right\rangle\simeq t^{\alpha}$, where the anomalous diffusion exponent $\alpha$ is less than one for subdiffusive processes and greater than one for superdiffusive processes, see e.g. \cite{metzler report}. The comb model (\ref{diffusion like eq on a comb0}) for $\gamma(t)=\eta(t)=\frac{t^{-\mu}}{\Gamma(1-\mu)}$ ($0<\mu<1$) yields the fractional comb model considered in \cite{iomin,mendez}, where the fractional derivatives appear in the form of the Caputo time fractional derivative
\begin{eqnarray}\label{Caputo_derivative}
{_{C}D_{t}^{q}}f(t)=\frac{1}{\Gamma(1-q)}\int_{0}^{t}dt'\,(t-t')^{-q}\frac{d}{dt'}W(t')
\end{eqnarray}
and the Riemman-Liouville fractional integral
\begin{eqnarray}\label{RLintegral}
{_{RL}I_{t}^{q}}f(t)=\frac{1}{\Gamma(q)}\int_{0}^{t}dt'\,(t-t')^{q-1}f(t').
\end{eqnarray}

This paper is organized as follows. In Section 2 we give analytical results for generalized fractal comb model. Different memory kernels are used and anomalous superdiffusion is observed. The connection between fractal structure of fingers and the Riesz fractional derivative is presented in Section 3. Summary is given in Section 4. At the end of the paper an additional material necessary for understanding of the main text is presented in Appendices. These relate to definitions and properties of the Mittag-Leffler, Fox $H$ and Weierstrass functions. Calculations of the PDFs and fractional moments are also presented in \ref{app solution}. Here we stress that we perform exact analytical analysis throughout the whole manuscript.

\section{Model formulation and solution}

At the first step of the present analysis let us understand the role of the $\delta(y)$ function in the highly inhomogeneous diffusion coefficients in Eqs.~(\ref{classical comb}) and~(\ref{diffusion like eq on a comb0}). One should recognize that the singularity of the $x$ component of the diffusion coefficient results from the Liouville equation; it is the intrinsic transport property of the comb models~(\ref{classical comb}) and~(\ref{diffusion like eq on a comb0}). Note that this singularity of the diffusion coefficient relates to a non-zero flux along the $x$ coordinates. Let us consider the Liouville equation
\begin{eqnarray}
\label{liouville eq}
\frac{\partial}{\partial t}P+\mathrm{div}\,\mathbf{j}=0,
\end{eqnarray}
where the two dimensional current $\mathbf{j}=(j_x,\,j_y)=\left(-\delta(y)\frac{\partial}{\partial x}P,\,-\frac{\partial}{\partial y}P\right)$ describes Markov processes in Eq.~(\ref{classical comb}). However, diffusion in both the backbone and fingers can be in general non-Markovian processes, which is reflected in Eq.~(\ref{diffusion like eq on a comb0}). Moreover the fingers can be inhomogeneously distributed as occurs in dendritic spines, where the spines are randomly (rather than uniformly) distributed \cite{fedotov}. In this case the two-dimensional current reads
\begin{eqnarray}\label{JxJy}
& j_x=-\mathcal{D}_x\delta(y)\int dt' \eta'(t-t')\frac{\partial}{\partial x}P(x,y,t'),\\
& j_y=-\mathcal{D}_y\int dx'dt'\gamma'(t-t')\rho(x-x')\frac{\partial}{\partial y}P(x',y,t').
\end{eqnarray}
Eq.~(\ref{liouville eq}) together with Eqs.~(6) and (7), can be regarded as the two-dimensional non-Markovian master equation. Integrating Eq.~(\ref{liouville eq}) over $y$ from $-\epsilon/2$ to $\epsilon/2$:
$\int_{-\epsilon/2}^{\epsilon/2}dy\dots $, one obtains for the l.h.s. of 
the equation, after application of the middle point theorem, 
$\epsilon\frac{\partial}{\partial t}P(x,y=0,t)$,
which is exact in the limit $\epsilon\rightarrow 0$.
This term can be neglected in the limit $\epsilon\rightarrow 0$.
Considering integration of the r.h.s. of the equation, we obtain that the term responsible for the transport in the $y$ direction reads from Eq.~(7)
$$\int dt'dx'\gamma'(t-t')\rho(x-x')\frac{\partial}{\partial y}
\Big[P(x',y,t')\big|_{y=\epsilon/2}-
  P(x,y,t')\big|_{y=-\epsilon/2}\Big] \, .$$
This corresponds to the two outgoing fluxes from the backbone in the $\pm 
y$ directions: $F_{y}(y=+0)+F_{y}(y=-0)$.
The transport along the $x$ direction, after integration of Eq.~(6), is
$$\epsilon \mathcal{D}(y\rightarrow 0)\frac{\partial^{2}}{\partial x^{2}}\int dt'\eta'(t-t')P(x,y=0,t')=
F_{x}(x+\epsilon)+F_{x}(x-\epsilon)\, .$$
Here, we take a general diffusivity function in the $x$ direction $\mathcal{D}(y)$
(instead of $\mathcal{D}_x\delta(y)$ in Eq.~(\ref{liouville eq}) and~(6)).
It should be stressed that the second derivative over $x$,
presented in the form
$\epsilon\frac{\partial^{2}}{\partial x^{2}}P=\left[\frac{\partial}{\partial x}P(x+\epsilon/2)-\frac{\partial}{\partial x}P(x-\epsilon/2)\right]$ as $\epsilon\rightarrow0$,
ensures both incoming and outgoing fluxes for $F_{x}$ along the $x$
direction at a point $x$. After integration over $y\in 
[-\epsilon,+\epsilon]$, the Liouville equation is a kind of the 
Kirchhoff's law: $F_{x}(+)+F_{x}(-)+F_{y}(+)+F_{y}(-)=0$
for each point $x$ and at $y=0$.
Since $j_x\neq 0$, outgoing fluxes are not zero, the flux 
$F_{x}\equiv F_{x}(+)+F_{x}(-)$ 
has to be nonzero as well: $F_{x}(\pm)\neq 0$.
Therefore, $\epsilon D(y\rightarrow 0)\neq 0$.
Taking the diffusion coefficient in the form $\mathcal{D}(y)=\frac{1}{\pi}\frac{\epsilon\mathcal{D}_{x}}{y^2+\epsilon^2}$, one obtains
in the limit $\epsilon\rightarrow 0$ a nonzero flux $F_{x}$ with
$\mathcal{D}(y)=\mathcal{D}_x\delta(y)$, which is the diffusion coefficient in the $x$ 
direction in Eqs.~(\ref{diffusion like eq on a comb0}),~(\ref{liouville eq}) and~(6). The relations between kernels $\gamma(t)$, $\eta(t)$ and $\gamma'(t)$, and $\eta'(t)$ in Eqs.~(\ref{diffusion like eq on a comb0}) and~(\ref{liouville eq}),~(6) and~(7) can be established in the Laplace space. Namely, performing the variable change in the Laplace space $\mathcal{L}[\gamma(t)]=\mathcal{L}[\gamma'(t)]$ and $\mathcal{L}[\eta(t)]=\mathcal{L}[\eta'(t)]/\mathcal{L}[\gamma'(t)]$ one arrives at Eq.~(\ref{diffusion like eq on a comb0}).

Presenting the last term in Eq.~(\ref{diffusion like eq on a comb0}) in the Fourier inversion form, Eq.~(\ref{diffusion like eq on a comb0}) reads
\begin{eqnarray}\label{diffusion like eq on a comb}
\int_{0}^{t}dt'\,\gamma(t-t')\frac{\partial}{\partial t'}P(x,y,t')
&=\mathcal{D}_{x}\delta(y)\int_{0}^{t}dt'\,\eta(t-t')\frac{\partial^{2}}{\partial x^{2}}P(x,y,t')\nonumber\\&+\mathcal{D}_{y}\mathcal{F}_{\kappa_{x}}^{-1}\left[|\kappa_{x}|^{1-\nu}\frac{\partial^{2}}{\partial y^{2}}\tilde{P}(\kappa_x,y,t)\right],
\end{eqnarray}
where $\rho(x)\sim|x|^{\nu-2}$ is used. Therefore, Eq.~(\ref{diffusion like eq on a comb}) can be presented by means of the Riesz space fractional derivative\footnote{The Riesz fractional derivative of order $\alpha$ ($0<\alpha\leq2$) is given as a pseudo-differential operator with the Fourier symbol
$-|\kappa|^\alpha$, $\kappa\in \mathrm{R}$ \cite{feller,samko}, i.e., $\frac{\partial^\alpha}{\partial|x|^\alpha}f(x)=\mathcal{F}^{-1}\left[-|\kappa|^\alpha\tilde{F}(\kappa)\right](x)$.} $\frac{\partial^{1-\nu}}{\partial |x|^{1-\nu}}$ of order $0<1-\nu<1$ \cite{samko}. This fractional derivative appears as a result of presenting the fingers density $|x|^{\nu-1}$ in the form of the Fourier transform\footnote{Originally the finger term reads $\mathcal{D}_y|x|^{\nu-1}\frac{\partial^{2}}{\partial y^2}P(x,y,t)$, see Ref.~\cite{iomin2}.}. This natural generalization of Eq.~(\ref{diffusion like eq on a comb}) establishes a relation between the fractal geometry of the medium and fractional integro-differentiation, where the reciprocal fractional density $|\kappa_x|^{1-\nu}$ leads to the fractional Riesz derivative of the order $0<1-\nu<1$. We also admit here that for $\nu=1$ ($\rho(x)=\delta(x)$) we call Eq.~(\ref{diffusion like eq on a comb0}) and Eq.~(\ref{diffusion like eq on a comb}) ``continuous'' comb, while for $\nu<1$ it is ``fractal'' comb model.

\subsection{PDF and $q$-th moment along the backbone}

To understand the properties of anomalous diffusion, one calculates the MSD. However, the MSD can diverge for L\'evy processes. In this case one calculates a fractional $q$-th moment, which is obtained here.

The Fourier-Laplace transforms of Eq.~(\ref{diffusion like eq on a comb}) yield
\begin{eqnarray}\label{Levy processes on comb}
\hat{\gamma}(s)\left[s\tilde{\hat{P}}(\kappa_{x},\kappa_{y},s)-\tilde{P}(\kappa_{x},\kappa_{y},t=0)\right]
&=-\mathcal{D}_{x}\kappa_{x}^{2}\hat{\eta}(s)\tilde{\hat{P}}(\kappa_{x},y=0,s)\nonumber\\&-\mathcal{D}_{y}|\kappa_{x}|^{1-\nu}\kappa_{y}^{2}\tilde{\hat{P}}(\kappa_{x},\kappa_{y},s),
\end{eqnarray}
where $\tilde{\hat{P}}(\kappa_{x},y,s)=\mathcal{F}_{x}\left[\mathcal{L}\left[P(x,y,t)\right]\right]$ and $\tilde{\hat{P}}(\kappa_{x},\kappa_{y},s)=\mathcal{F}_{y}\left[\tilde{\hat{P}}(\kappa_{x},y,s)\right]$. Performing the inverse Fourier transform of $\tilde{\hat{P}}(\kappa_x,\kappa_y,s)$ with respect to $\kappa_{y}$, one finds $\tilde{\hat{P}}(\kappa_{x},y,s)$, from where $\tilde{\hat{P}}(\kappa_{x},y=0,s)$ reads
\begin{eqnarray}\label{diffusion like eq Levy y=0}
\tilde{\hat{P}}(\kappa_{x},y=0,s)=\left.\frac{1}{s}\sqrt{\frac{s\hat{\gamma}(s)}{4\mathcal{D}_{y}}}|\kappa_{x}|^{\frac{\nu-1}{2}}\right/\left[1+\frac{1}{s}\sqrt{\frac{s\hat{\gamma}(s)}{4\mathcal{D}_{y}}}\frac{\mathcal{D}_{x}\hat{\eta}(s)}{\hat{\gamma}(s)}|\kappa_{x}|^{\frac{3+\nu}{2}}\right].
\end{eqnarray}
Here we use the initial condition $\tilde{P}(\kappa_x,\kappa_y,t=0)=1$. Substituting Eq.~(\ref{diffusion like eq Levy y=0}) in Eq.~(\ref{Levy processes on comb}), one obtains
\begin{eqnarray}\label{diffusion like eq Levy FFL}
\tilde{\hat{P}}(\kappa_{x},\kappa_{y},s)=
\frac{s\hat{\gamma}(s)\hat{\xi}(s)}{\left(s\hat{\gamma}(s)+\mathcal{D}_{y}|\kappa_{x}|^{1-\nu}\kappa_{y}^{2}\right)\left(s\hat{\xi}(s)+\frac{\mathcal{D}_{x}}{2\sqrt{\mathcal{D}_{y}}}|\kappa_{x}|^{\frac{3+\nu}{2}}\right)}.
\end{eqnarray}
Taking  $\kappa_y=0$ in Eq.~(\ref{diffusion like eq Levy FFL}), which eventually leads to the reduced PDF $p_{1}(x,t)=\int_{-\infty}^{\infty}dy\,P(x,y,t)$, one obtains  the latter in the form
\begin{eqnarray}\label{p1 general Levy}
\tilde{\hat{p}}_{1}(\kappa_{x},s)=\frac{\hat{\xi}(s)}{s\hat{\xi}(s)+\frac{\mathcal{D}_{x}}{2\sqrt{\mathcal{D}_{y}}}|\kappa_{x}|^{\frac{3+\nu}{2}}},
\end{eqnarray}
where $\tilde{\hat{p}}_{1}(\kappa_{x},s)=\mathcal{F}_{x}\left[\mathcal{L}\left[p_{1}(x,t)\right]\right]$, and
\begin{eqnarray}\label{xi(s)} \hat{\xi}(s)=\frac{1}{\hat{\eta}(s)}\sqrt{\frac{\hat{\gamma}(s)}{s}}.
\end{eqnarray}
Equation (\ref{p1 general Levy}) corresponds to the fractional diffusion equation
for the reduced distribution $p_{1}(x,t)$, which describes both L\'evy flights with traps and subdiffusion,
\begin{eqnarray}
\label{diffusion-like eq memory Levy}
\int_0^t dt'\,\xi(t-t')\frac{\partial}{\partial t'}p_{1}(x,t')=\frac{\mathcal{D}_{x}}{2\sqrt{\mathcal{D}_{y}}}
\frac{\partial^{\alpha}}{\partial |x|^{\alpha}}p_{1}(x,t).
\end{eqnarray}
Here the Riesz space fractional derivative is of order $\alpha=\frac{3+\nu}{2}\leq2$, while integro-differentiation with respect to time is presented in the Caputo form. 

Introducing a waiting times PDF $\psi(t)$, which in the Laplace space is given by $\hat{\psi}(s)=\left(1+s\hat{\xi}(s)\right)^{-1}$ \cite{sandev draft}, one obtains the relation $\hat{\xi}(s)=\frac{1-\hat{\psi}(s)}{s\hat{\psi}(s)}$. For example, in the Markov case, when 
$\psi(t)=\frac{1}{\tau}e^{-t/\tau}$, the trap kernel is a $\delta$ function and the l.h.s. of Eq.~(\ref{diffusion-like eq memory Levy}) reduces to the standard time derivative $\frac{\partial}{\partial t}p_1(x,t)$. A subdiffusive case, when $\psi(t)=\frac{1}{1+(t/\tau)^{\beta}}$, yields \cite{IS2012} $\hat{\xi}^{-1}(s)=s^{1-\beta}$. Then the l.h.s. of Eq.~(\ref{diffusion-like eq memory Levy}) corresponds to the Caputo fractional derivative of the order of $\beta$, defined in Eq.~(\ref{Caputo_derivative}). Therefore, the power-law tail of the kernel $\xi(t)$ determines the Caputo fractional derivative (\ref{Caputo_derivative}).

It is worth mentioning that the solution of Eq.~(\ref{diffusion like eq on a comb}) in the Fourier-Laplace space $(\kappa_{x},s)$ can be written as
\begin{eqnarray}\label{P kx s}
\tilde{\hat{P}}(\kappa_{x},y,s)=\exp\left(-\sqrt{\frac{s\tilde{\hat{g}}(\kappa_{x},s)}{\mathcal{D}_{y}}}|y|\right)\tilde{\hat{f}}(\kappa_{x},s),
\end{eqnarray}
where $f(x,t)$ and $g(x,t)$ are functions standing for the derivation. We find that $\tilde{\hat{g}}(\kappa_{x},s)=\hat{\gamma}(s)|\kappa_{x}|^{\nu-1}$ and $\tilde{\hat{f}}(\kappa_{x},s)$ is given by Eq.~(\ref{diffusion like eq Levy y=0}), from where the Fourier transform in respect to $y$, gives the same expression for $\tilde{\hat{P}}(\kappa_{x},\kappa_{y},s)$ as in Eq.~(\ref{diffusion like eq Levy FFL}).

The $q$-th fractional moments can be analyzed for various forms of the kernels $\gamma(t)$ and $\eta(t)$  
\begin{eqnarray}\label{fractional moments}
\left\langle|x(t)|^{q}\right\rangle=2\int_{0}^{\infty}dx\,x^{q}p_{1}(x,t),
\end{eqnarray}
where $0<q<\alpha<2$. One considers the $q$-th fractional moments with $q<\alpha$, since the MSD for L\'{e}vy processes governed by equation (\ref{diffusion-like eq memory Levy}) does not exist. Therefore, instead of the MSD one can analyze its analogue related to the $q$-th moment, $\left\langle|x(t)|^{q}\right\rangle^{2/q}$ \cite{metzler report}. From relation (\ref{p1 general Levy}) one obtains
\begin{eqnarray}\label{fractional moments laplace}
\left\langle|x(t)|^{q}\right\rangle=C_{\alpha}(q)\mathcal{L}^{-1}\left[\frac{1}{s\left(s\hat{\xi}(s)\right)^{q/\alpha}}\right],
\end{eqnarray}
where
\begin{eqnarray}\label{Cq}
C_{\alpha}(q)=\frac{4}{\alpha}\left(\frac{\mathcal{D}_{x}}{2\sqrt{\mathcal{D}_{y}}}\right)^{q/\alpha}\frac{\Gamma(q)\Gamma\left(1+q/\alpha\right)\Gamma\left(-q/\alpha\right)}{\Gamma(q/2)\Gamma(-q/2)}.
\end{eqnarray}
The case with $\nu=1$, i.e., $\alpha=2$ (continuous comb), yields
\begin{eqnarray}\label{fractional moments laplace nu=1}
\left\langle|x(t)|^{q}\right\rangle=\Gamma(q+1)\left(\frac{\mathcal{D}_{x}}{2\sqrt{\mathcal{D}_{y}}}\right)^{q/2}\mathcal{L}^{-1}\left[\frac{1}{s\left(s\hat{\xi}(s)\right)^{q/2}}\right],
\end{eqnarray}
from where for $q=2$ we recover the result for the MSD \cite{joint MMNP}
\begin{eqnarray}\label{fractional moments laplace nu=1 msd}
\left\langle x^{2}(t)\right\rangle=\frac{\mathcal{D}_{x}}{\sqrt{\mathcal{D}_{y}}}\mathcal{L}^{-1}\left[\frac{1}{s^{2}\hat{\xi}(s)}\right].
\end{eqnarray}
Now setting different functional behaviors of the kernels $\gamma(t)$ and $\eta(t)$,
one can observe various diffusion regimes, along both the $x$ and $y$ directions.

\subsection{Special case I: L\'{e}vy distribution}

When $\gamma(t)=\delta(t)$, i.e., $\hat{\gamma}(s)=1$, and $\eta(t)=\frac{t^{-1/2}}{\Gamma(1/2)}$, i.e., $\hat{\eta}(s)=s^{-1/2}$, which means $\hat{\xi}(s)=1$, we obtain the Markovian transport equation for superdiffusion along the backbone
\begin{eqnarray}
\label{diffusion-like eq delta memory Levy}
\frac{\partial}{\partial t}p_{1}(x,t)=\frac{\mathcal{D}_{x}}{2\sqrt{\mathcal{D}_{y}}}
\frac{\partial^{\alpha}}{\partial |x|^{\alpha}}p_{1}(x,t).
\end{eqnarray}
Taking the initial condition $p_{1}(x,0+)=\delta(x)$ and the boundary conditions $p_{1}(\pm\infty,t)=\frac{\partial}{\partial x}p_{1}(\pm\infty,t)=0$ (see \ref{app solution}), one obtains the solution of Eq.~(\ref{diffusion-like eq delta memory Levy})
\begin{eqnarray}
\label{diffusion-like eq delta memory Levy sol}
p_{1}(x,t)=\frac{1}{\alpha|x|} H_{3,3}^{2,1}\left[\left.\frac{|x|}{\left(\frac{\mathcal{D}_{x}}{2\sqrt{\mathcal{D}_{y}}}t\right)^{1/\alpha}}
\right|\left.\begin{array}{l}
    (1,\frac{1}{\alpha}),(1,\frac{1}{\alpha}),(1,\frac{1}{2})\\
    (1,1),(1,\frac{1}{\alpha}),(1,\frac{1}{2})
  \end{array}\right.\right],
\end{eqnarray}
where $H_{p,q}^{m,n}\left[z\left|\begin{array}{l}
    (a_p,A_p)\\
    (b_q,B_q)
  \end{array}\right.\right]$ is the Fox $H$-function \cite{saxena book} (see also a brief introduction in \ref{WF}).
  
Therefore,  the $q$-th moment reads (see calculations in \ref{app solution})
\begin{eqnarray}\label{fractional moments laplace delta}
\left\langle|x(t)|^{q}\right\rangle=C_{\alpha}(q)\frac{t^{q/\alpha}}{\Gamma\left(1+q/\alpha\right)},
\end{eqnarray}
where $C_{\alpha}(q)$ is defined in Eq.~(\ref{Cq}). From Eq.~(\ref{fractional moments laplace delta}) one obtains $\left\langle|x(t)|^{q}\right\rangle^{2/q}\simeq t^{4/(3+\nu)}$ that corresponds to superdiffusion (L\'evy flights \cite{west}) since $0<\nu<1$. The same superdiffusive behavior is observed when $s^{-1/2}\sqrt{\hat{\gamma}(s)}=\hat{\eta}(s)$, which means $\hat{\xi}(s)=1$. Note that Eq.~(\ref{Levy processes on comb}) describes a typical competition between long rests and long jumps \cite{metzler report}. Contrary to the case described in Refs.~\cite{iomin2,mendez}, in the present analysis, superdiffusion can be dominant not only due to the fractional (power-law) distribution of the fingers with $0<\nu<1$, but also due to the specific choice of the time kernels $\eta(t)$ and $\gamma(t)$.

\subsection{Special case II: Competition between long rests and L\'{e}vy flights}

Now we consider the power-law memory kernels in the form $\gamma(t)=\eta(t)=\frac{t^{-\mu}}{\Gamma(1-\mu)}$, $0<\mu<1$. From Eq.~(\ref{xi(s)}) we find $\hat{\xi}(s)=s^{-\mu/2}$, which yields in the time domain that $\xi(t)=\frac{t^{-(1-\mu/2)}}{\Gamma(\mu/2)}$. Therefore, the space-time fractional diffusion equation for the reduced PDF $p_{1}(x,t)$ is a non-Markovian trasport equation for superdiffusion along the backbone
\begin{eqnarray}\label{diffusion-like eq two power law memory Levy}
{_{C}}D_{t}^{1-\mu/2}p_{1}(x,t)=\frac{\mathcal{D}_{x}}{2\sqrt{\mathcal{D}_{y}}}
\frac{\partial^{\alpha}}{\partial |x|^{\alpha}}p_{1}(x,t),
\end{eqnarray} 
where ${_{C}}D_{t}^{1-\mu/2}$ is the Caputo time fractional derivative (\ref{Caputo_derivative}) of order $1/2<1-\mu/2<1$, and $\frac{\partial^{\alpha}}{\partial |x|^{\alpha}}$ is the Riesz space fractional derivative of order $\alpha=\frac{3+\nu}{2}$. The initial condition is $p_{1}(x,0+)=\delta(x)$, and the boundary conditions are defined at infinities $p_{1}(\pm\infty,t)=\frac{\partial}{\partial x}p_{1}(\pm\infty,t)=0$. Taking into account the initial and the boundary conditions, one obtains the solution of Eq.~(\ref{diffusion-like eq two power law memory Levy}) in terms of the Fox $H$-function (see \ref{app solution}, Eq.~(\ref{special case}))
\begin{eqnarray}\label{sol two power law memory Levy} p_{1}(x,t)=\frac{1}{\alpha|x|}H_{3,3}^{2,1}\left[\left.\frac{|x|}{\left(\frac{\mathcal{D}_{x}}{2\sqrt{\mathcal{D}_{y}}}t^{1-\mu/2}\right)^{1/\alpha}}
\right|\left.\begin{array}{l}
    (1,\frac{1}{\alpha}),(1,\frac{1-\mu/2}{\alpha}),(1,\frac{1}{2})\\
    (1,1),(1,\frac{1}{\alpha}),(1,\frac{1}{2})
  \end{array}\right.\right].
\end{eqnarray} 

Repeating the calculation of the fractional $q$-th moment in Eq.~(\ref{moments case}),
one obtains 
\begin{eqnarray}\label{fractional moments laplace two power law}
\left\langle|x(t)|^{q}\right\rangle=C_{\alpha}(q)\frac{t^{\frac{2-\mu}{2\alpha}q}}{\Gamma\left(1+\frac{2-\mu}{2\alpha}q\right)},
\end{eqnarray}
which also yields $\left\langle|x(t)|^{q}\right\rangle^{2/q}\simeq t^{(2-\mu)/\alpha}$. One concludes here that superdiffusion appears for $2\mu+\nu<1$, and subdiffusion takes place for $2\mu+\nu>1$. These effects result from the combination of the memory kernels that eventually leads to the competition between long rests and long jumps. Note that in the limit case of $\nu=1$, there is subdiffusion with the correct MSD $\left\langle x^{2}(t)\right\rangle\simeq t^{1-\mu/2}$ \cite{metzler report,sandev jpa2011}.

\subsection{Special case III: Distributed order memory kernels}

Note that there are many choices of the memory kernels that can lead to more specific situations.  For example, as it is shown in Refs.~\cite{chechkin2,mainardi_book,sandev draft},  distributed order memory kernels can lead to a strong anomaly in fractional kinetics
like ultra-slow diffusion, where  for example the Sinai diffusion \cite{Sinai} is one of  the best-known realizations of anomalous kinetics. 

Let us consider the distributed order memory kernel of the form \cite{chechkin2,chechkin,kochubei,mainardi_book}
\begin{eqnarray}\label{distibuted memory kernel gamma}
\gamma(t)=\int_{0}^{1}d\mu\,\frac{t^{-\mu}}{\Gamma(1-\mu)},
\end{eqnarray}
which yields $\gamma(s)=\frac{s-1}{s\log(s)}$ \cite{chechkin2,mainardi_book}, and for $\eta(t)=\delta(t)$ one obtains $\xi(s)=\frac{1}{s}\sqrt{\frac{s-1}{\log{s}}}$. For the calculation of the $q$-th moment, it is convenient to use here the Tauberian theorem \cite{feller}, which states that for a slowly varying function $L(t)$ at infinity, i.e., $\lim_{t\rightarrow\infty}\frac{L(at)}{L(t)}=1$, $a>0$, if
$\hat{R}(s)\simeq s^{-\rho}L\left(\frac{1}{s}\right)$, $s\rightarrow0$, $\rho\geq0,$
then
$r(t)=\mathcal{L}^{-1}\left[\hat{R}(s)\right]\simeq
\frac{1}{\Gamma(\rho)}t^{\rho-1}L(t)$, $t\rightarrow\infty$. Therefore, applying the Tauberian theorem, one obtains the behavior of the fractional $q$-th moments in the long time limit
\begin{eqnarray}\label{MSDx distributed 1}
\fl \left\langle |x(t)|^{q}\right\rangle= C_{\alpha}(q)\mathcal{L}^{-1}\left[\frac{1}{s}\left(\frac{\log{s}}{s-1}\right)^{\frac{q}{2\alpha}}\right]\simeq C_{\alpha}(q)\mathcal{L}^{-1}\left[\frac{1}{s}\left(\log{\frac{1}{s}}\right)^{\frac{2q}{\alpha}}\right]\simeq C_{\alpha}(q)\log^{\frac{q}{2\alpha}}t,
\end{eqnarray}
which yields $\left\langle |x(t)|^{q}\right\rangle^{2/q}\simeq\log^{\frac{1}{\alpha}}t$. This result also contains the correct limit of the continuous comb with $\nu=1$ ($\alpha=2$),
when the MSD reads $\left\langle x^{2}(t)\right\rangle\simeq\frac{\mathcal{D}_{x}}{\sqrt{\mathcal{D}_{y}}}\log^{1/2}t$ \cite{joint MMNP}. It should be stressed that ultra-slow diffusion takes place here even in the presence of the L\'evy flights. However the latter affects only the power of the logarithm, since ultra-slow diffusion is the robust process with respect to  the inhomogeneous distribution of the fingers.     

For a more general distributed order memory kernel of the form \cite{chechkin}
\begin{eqnarray}\label{distibuted memory kernel gamma nu}
\gamma(t)=\int_{0}^{1}d\mu\,\lambda\mu^{\lambda-1}\frac{t^{-\mu}}{\Gamma(1-\mu)},
\end{eqnarray}
where $\lambda>0$, one obtains for the long time limit $\gamma(s)\simeq\frac{\Gamma(1+\lambda)}{s\log^{\lambda}\frac{1}{s}}$, and for $\eta(t)=\delta(t)$ the $q$-th moment reads
\begin{eqnarray}\label{MSDx distributed 2}
\left\langle |x(t)|^{q}\right\rangle\simeq C_{\alpha}(q)\mathcal{L}^{-1}\left[\frac{1}{s}\left(\frac{\log^{\lambda}{\frac{1}{s}}}{\Gamma(1+\lambda)}\right)^{\frac{q}{2\alpha}}\right]\simeq C_{\alpha}(q)\left(\frac{\log^{\lambda}t}{\Gamma(1+\nu)}\right)^{\frac{q}{2\alpha}}.
\end{eqnarray}
This $q$-th moment behavior eventually yields $\left\langle |x(t)|^{q}\right\rangle^{2/q}\simeq\left(\frac{\log^{\lambda}t}{\Gamma(1+\nu)}\right)^{1/\alpha}$, which also contains the limiting case of the continuous comb with the MSD $\left\langle x^{2}(t)\right\rangle\simeq\frac{\mathcal{D}_{x}}{\sqrt{\mathcal{D}_{y}}}\frac{\log^{\lambda/2}t}{\sqrt{\Gamma(1+\lambda)}}$ \cite{joint MMNP}.

\subsection{Diffusion along fingers}

One easily finds that the solution (\ref{diffusion like eq Levy FFL}) does not describe diffusion in the $y$ direction. Indeed, it follows from Eq.~(\ref{diffusion like eq Levy FFL}) that
\begin{eqnarray}\label{p2 general}
\tilde{\hat{p}}_{2}(\kappa_{y},s)=\frac{1}{s},
\end{eqnarray}
where $p_{2}(y,t)=\int_{-\infty}^{\infty}dx\,P(x,y,t)$, which means that $p_{2}(y,t)=\delta(y)$, from where one obtains that the MSD along the $y$-direction is equal to zero. However diffusion in the $y$ direction does take place with the diffusivity $\mathcal{D}_y$. To resolve this paradox, one should understand that the MSD is obtained by averaging over the total volume, which yields zero power of the set: $\lim_{L\to\infty}\frac{1}{L}\int_{0}^{L}dx\,x^{\nu-1}\sim \lim_{L\to\infty}L^{\nu-1}=0$.
To obtain a finite result, one has to average over the fractal volume $L^{\nu}$.
Therefore, the Fourier inversion over the fractal measure $|\kappa_x|^{\nu-1}d\kappa_x$ yields for the MSD
\begin{eqnarray}
&\left\langle y^{2}(t)\right\rangle=\mathcal{L}^{-1}\left.\left[-\frac{\partial^{2}}{\partial \kappa_{y}^{2}}\tilde{\hat{P}}(\kappa_{x},\kappa_{y},s)|\kappa_x|^{\nu-1}\right]\right|_{\kappa_x=0,\kappa_{y}=0}\nonumber\\&=\mathcal{L}^{-1}\left.\left[\frac{2\mathcal{D}_{y}s\hat{\gamma}(s)-6\mathcal{D}_{y}^{2}|\kappa_x|^{1-\nu}\kappa_{y}^{2}}{\left(s\hat{\gamma}(s)+\mathcal{D}_{y}|\kappa_{x}|^{1-\nu}\kappa_{y}^{2}\right)^{3}}\cdot\frac{s\hat{\gamma}(s)\hat{\xi}(s)}{s\hat{\xi}(s)+\frac{\mathcal{D}_{x}}{2\sqrt{\mathcal{D}_{y}}}|\kappa_{x}|^{\frac{3+\nu}{2}}}\right]\right|_{\kappa_x=0,\kappa_{y}=0}\nonumber\\ &=2\mathcal{D}_{y}\mathcal{L}^{-1}\left.\left[\frac{1}{s^{2}\hat{\gamma}(s)}\cdot\frac{s\hat{\xi}(s)}{s\hat{\xi}(s)+\frac{\mathcal{D}_{x}}{2\sqrt{\mathcal{D}_{y}}}|\kappa_{x}|^{\frac{3+\nu}{2}}}\right]\right|_{\kappa_{x}=0}=2\mathcal{D}_{y}\mathcal{L}^{-1}\left[\frac{1}{s^{2}\hat{\gamma}(s)}\right],\nonumber\\
\end{eqnarray}
where $\tilde{\hat{P}}(\kappa_{x},\kappa_{y},s)$ is given by Eq.~(\ref{diffusion like eq Levy FFL}). This result is the same as the one obtained for the generalized continuous comb model $\nu=1$ \cite{joint MMNP}, which follows from Eq.~(\ref{diffusion like eq Levy FFL}) for $\nu=1$. We finally note that for the various forms of the memory kernel $\gamma(t)$ one can find different diffusive regimes along the fingers, such as anomalous and ultraslow diffusion.

\section{Fractal structure of fingers and the Weierstrass function}

\subsection{General solution of the problem}

Let us rewrite the last term in Eq.~(\ref{diffusion like eq on a comb0}) in the form of the convolution with the Weierstrass function in the Fourier $\kappa_x$ space. This reads
\begin{eqnarray}\label{eq in FF space}
\mathcal{D}_{y}\frac{\partial^{2}}{\partial y^{2}}\frac{1}{2\pi}\int_{-\infty}^{\infty}d\kappa_{x}{'}\Psi\left(\kappa_{x}-\kappa_{x}{'}\right)\tilde{P}(\kappa_{x}{'},{y},t).
\end{eqnarray}
Here $\Psi\left(\kappa_{x}-\kappa_{x}{'}\right)$ is the
Weierstrass function \cite{berry,sandev iomin kantz} with the scaling property
\begin{eqnarray}\label{Weierstrass3}
\Psi(z/l)\simeq\frac{l}{b}\Psi(z),
\end{eqnarray}
which, for example, can be defined by the procedure suggested in \ref{WF}.

This scaling property leads to the power-law asymptotic behavior of the 
Weierstrass function $\Psi(z)\sim\frac{1}{z^{1-\bar{\nu}}}$, where
$\bar{\nu}=\log{b}/\log{l}$, with the fractal dimension $0<\bar{\nu}<1$. Therefore, the term in Eq.~(\ref{eq in FF space}) can be presented in the form of the Riesz fractional integral in the reciprocal Fourier space
\begin{eqnarray}\label{eq in Fx}
\mathcal{D}_{y}\frac{1}{2\pi}\frac{\partial^{2}}{\partial y^{2}}\int_{-\infty}^{\infty}d\kappa_{x}{'}\frac{\tilde{P}(\kappa_{x}{'},y,t)}{|\kappa_{x}-\kappa_{x}{'}|^{1-\bar{\nu}}}.
\end{eqnarray}
Applying the inverse Fourier transform in respect to $\kappa_x$, and changing the order of integration, one obtains
\begin{eqnarray}\label{eq in Fx}
\fl\mathcal{D}_{y}\frac{1}{2\pi}\frac{\partial^{2}}{\partial y^{2}}\mathcal{F}_{\kappa_x}^{-1}\left[\int_{-\infty}^{\infty}d\kappa_{x}{'}\tilde{P}(\kappa_{x}{'},y,t)\frac{1}{|\kappa_{x}-\kappa_{x}{'}|^{1-\bar{\nu}}}\right]=\mathcal{D}_{y}C_{\nu}|x|^{-\bar{\nu}}\frac{\partial^{2}}{\partial y^{2}}P(x,y,t),
\end{eqnarray}
where $C_{\bar{\nu}}=\Gamma(\bar{\nu})\cos{\frac{\bar{\nu}\pi}{2}}$. Thus, Eq.~(\ref{diffusion like eq on a comb0}) becomes
\begin{eqnarray}\label{diffusion like eq on a comb weierstrass final}
\int_{0}^{t}dt'\,\gamma(t-t')\frac{\partial}{\partial t'}P(x,y,t')
&=\mathcal{D}_{x}\delta(y)\int_{0}^{t}dt'\,\eta(t-t')\frac{\partial^{2}}{\partial x^{2}}P(x,y,t')\nonumber\\&+\mathcal{D}_{y}C_{\bar{\nu}}|x|^{-\bar{\nu}}\frac{\partial^{2}}{\partial y^{2}}P(x,y,t).
\end{eqnarray}
Note that in contrast to Eq.~(\ref{diffusion like eq on a comb0}), here the continuous comb model corresponds to the limit with $\bar{\nu}=0$. In this mean $\bar{\nu}$ is dual to $\nu$ with the  relation $\bar{\nu}+\nu=1$. Performing the Laplace transform, one obtains
\begin{eqnarray}\label{diffusion like eq on a comb weierstrass final laplace}
\fl\hat{\gamma}(s)\left[s\hat{P}(x,y,s)-\delta(x)\delta(y)\right]
=\mathcal{D}_{x}\delta(y)\hat{\eta}(s)\frac{\partial^{2}}{\partial x^{2}}\hat{P}(x,y,s)+\mathcal{D}_{y}C_{\bar{\nu}}|x|^{-\bar{\nu}}\frac{\partial^{2}}{\partial y^{2}}\hat{P}(x,y,s).\nonumber\\
\end{eqnarray}

By analogy with Eq.~(\ref{P kx s}), the solution of Eq.~(\ref{diffusion like eq on a comb weierstrass final laplace}) can be presented in the form
\begin{eqnarray}\label{solution representation}
\hat{P}(x,y,s)=\exp\left(-\sqrt{\frac{s\hat{g}(x,s)}{\mathcal{D}_{y}}}|y|\right)\hat{f}(x,s),
\end{eqnarray}
where $\hat{g}(x,s)$ is obtained from the condition that the second derivative of the exponential compensates the first term in the l.h.s. of Eq.~(\ref{diffusion like eq on a comb weierstrass final}). This reads
\begin{eqnarray}\label{solution g}
\hat{g}(x,s)=\frac{1}{C_{\bar{\nu}}}\hat{\gamma}(s)|x|^{\bar{\nu}},
\end{eqnarray}
and the solution $\hat{P}(x,y,s)$ becomes
\begin{eqnarray}\label{solution representation new}
\hat{P}(x,y,s)=\exp\left(-\sqrt{\frac{1}{C_{\bar{\nu}}}\frac{s\hat{\gamma}(s)}{\mathcal{D}_{y}}}|x|^{\bar{\nu}/2}|y|\right)\hat{f}(x,s).
\end{eqnarray}
From here we find that
\begin{eqnarray}\label{solution representation} \hat{p}_{1}(x,s)=\int_{-\infty}^{\infty}dy\,\hat{P}(x,y,s)=2\sqrt{\frac{\mathcal{D}_{y}}{s\hat{g}(x,s)}}\hat{f}(x,s), 
\end{eqnarray}
and
\begin{eqnarray}\label{solution representation y=0}
\hat{P}(x,y=0,s)=\hat{f}(x,s).
\end{eqnarray}

Integrating Eq.~(\ref{diffusion like eq on a comb weierstrass final laplace}) over $y$ and taking into account Eq.~(\ref{solution g}), one obtains the boundary value problem for the Green function $\hat{f}(x,s)$ with  zero boundary conditions at infinities
\begin{eqnarray}\label{f(x,s)2}
2C_{\bar{\nu}}^{1/2}\sqrt{\frac{\mathcal{D}_{y}s}{\hat{\gamma}(s)}}|x|^{-\bar{\nu}/2}\hat{f}(x,s)-\mathcal{D}_{x}\frac{\hat{\eta}(s)}{\hat{\gamma}(s)}\frac{\partial^{2}}{\partial x^{2}}\hat{f}(x,s)=\delta(x).
\end{eqnarray}
Follow the standard procedure, we consider the homogeneous part of the equation, which reads 
\begin{eqnarray}\label{f(x,s) green}
2C_{\bar{\nu}}^{1/2}\frac{\sqrt{\mathcal{D}_{y}s\hat{\gamma}(s)}}{\hat{\eta}(s)}|x|^{-\bar{\nu}/2}\hat{G}(x,s)=\mathcal{D}_{x}\frac{\partial^{2}}{\partial x^{2}}\hat{G}(x,s).
\end{eqnarray}

\subsection{Special case with $\gamma(t)=\eta(t)=\delta(t)$}

To be specific, we consider first a special case with $\hat{\gamma}(s)=\hat{\eta}(s)=1$.  Thus Eq.~(\ref{f(x,s) green}) reads
\begin{eqnarray}\label{f(x,s) green ex}
C_{\bar{\nu}}^{1/2}\frac{2\sqrt{\mathcal{D}_{y}}}{\mathcal{D}_{x}}s^{1/2}|x|^{-\bar{\nu}/2}\hat{G}(x,s)=\frac{\partial^{2}}{\partial x^{2}}\hat{G}(x,s).
\end{eqnarray}
It is symmetric with respect to $x\rightarrow -x$ and has a form of the Lommel differential equation $u''(x)-c^{2}x^{2\zeta-2}u(x)=0$ \cite{book integrals}. The solution is given in terms of the Bessel functions $u(x)=\sqrt{x}Z_{\frac{1}{2\zeta}}\left(\imath\frac{c}{\zeta}x^{\zeta}\right)$, where $Z_{\frac{1}{2\zeta}}(x)=C_{1}J_{\frac{1}{2\zeta}}(x)+C_{2}N_{\frac{1}{2\zeta}}(x)$. Here $J_{\frac{1}{2\zeta}}(x)$ is the Bessel function of the first kind and $N_{\frac{1}{2\zeta}}(x)$ is the Bessel function of the second kind (Neumann function). Therefore, the solution of Eq.~(\ref{f(x,s) green ex}) reads
\begin{eqnarray}\label{f(x,s) green ex sol}
\hat{G}(x,s)=\sqrt{x}Z_{\frac{2}{4-\bar{\nu}}}\left(i\,C_{\bar{\nu}}^{1/4}\frac{4}{4-\bar{\nu}}\sqrt{\frac{2\sqrt{\mathcal{D}_{y}}}{\mathcal{D}_{x}}}s^{1/4}x^{\frac{4-\bar{\nu}}{4}}\right).
\end{eqnarray}
Due to the zero boundary conditions, Green's function (\ref{f(x,s) green ex sol}) is given by the modified Bessel function (of the third kind) $K_{\frac{2}{4-\bar{\nu}}}(z)$, which can be expressed in terms of the Fox $H$-function as well (see relation (\ref{HK relation}))
\begin{eqnarray}\label{f(x,s) green ex sol K}
\hat{G}(x,s)&=\sqrt{x}K_{\frac{2}{4-\bar{\nu}}}\left(C_{\bar{\nu}}^{1/4}\frac{4}{4-\bar{\nu}}\sqrt{\frac{2\sqrt{\mathcal{D}_{y}}}{\mathcal{D}_{x}}}s^{1/4}x^{\frac{4-\bar{\nu}}{4}}\right)\nonumber\\&=\frac{\sqrt{x}}{2}H_{0,2}^{2,0}\left[\left.\frac{4C_{\bar{\nu}}^{1/2}}{(4-\bar{\nu})^{2}}\frac{2\sqrt{\mathcal{D}_{y}}}{\mathcal{D}_{x}}x^{\frac{4-\bar{\nu}}{2}}s^{1/2}\right|\left.\begin{array}{l} \\
(\frac{1}{4-\bar{\nu}},1),(-\frac{1}{4-\bar{\nu}},1)\end{array}\right.\right].
\end{eqnarray}

Considering the inhomogeneous Lommel Eq.~(\ref{f(x,s)2}), we use the solution $\hat{f}(|x|,s)=\mathcal{C}_{\bar{\nu}}(s)\hat{G}(|x|,s)=\mathcal{C}_{\bar{\nu}}(s)\hat{G}(y,s)$ obtained in Eq.~(\ref{f(x,s)2}), where $y=|x|$, and $\mathcal{C}_{\bar{\nu}}(s)$ is a function which depends on $s$,
\begin{eqnarray}\label{condition}
-2\mathcal{D}_{x}\frac{\partial}{\partial y}\hat{f}(y=0,s)=1.
\end{eqnarray}
Substituting Eq.~(\ref{f(x,s) green ex sol K}) in Eq.~(\ref{f(x,s)2}), and using relations (\ref{condition}) and (\ref{K series}), one obtains 
\begin{eqnarray}
\mathcal{C}_{\bar{\nu}}(s)=\frac{2}{4-\bar{\nu}}\frac{1}{\Gamma\left(\frac{2-\bar{\nu}}{4-\bar{\nu}}\right)\mathcal{D}_{x}}\left(C_{\bar{\nu}}^{1/2}\frac{4}{(4-\bar{\nu})^{2}}\frac{2\sqrt{\mathcal{D}_{y}}}{\mathcal{D}_{x}}\right)^{-\frac{1}{4-\bar{\nu}}}s^{-\frac{1}{2(4-\bar{\nu})}},
\end{eqnarray}
which yields the solution of Eq.~(\ref{f(x,s)2})
\begin{eqnarray}\label{f final ex}
\hat{f}(x,s)&=\frac{1}{4-\bar{\nu}}\frac{1}{\Gamma\left(\frac{2-\bar{\nu}}{4-\bar{\nu}}\right)\mathcal{D}_{x}}\left(C_{\bar{\nu}}^{1/2}\frac{4}{(4-\bar{\nu})^{2}}\frac{2\sqrt{\mathcal{D}_{y}}}{\mathcal{D}_{x}}\right)^{-\frac{1}{4-\bar{\nu}}}s^{-\frac{1}{2(4-\bar{\nu})}}|x|^{1/2}\nonumber\\&\times H_{0,2}^{2,0}\left[\left.\frac{4C_{\bar{\nu}}^{1/2}}{(4-\bar{\nu})^{2}}\frac{2\sqrt{\mathcal{D}_{y}}}{\mathcal{D}_{x}}|x|^{\frac{4-\bar{\nu}}{2}}s^{1/2}\right|\left.\begin{array}{l} \\
(\frac{1}{4-\bar{\nu}},1),(-\frac{1}{4-\bar{\nu}},1)\end{array}\right.\right].
\end{eqnarray}
From relations (\ref{solution representation}) and (\ref{H_laplace}), one finds
the solution for the reduced PDF $p_1(x,t)$\footnote{One can easily check from relations (\ref{H property1}) and (\ref{integral of H}) that $p_{1}(x,t)$ is normalized $\int_{-\infty}^{\infty}dx\,p_{1}(x,t)=1$.}
\begin{eqnarray}\label{p1(x,t) final ex}
p_{1}(x,t)&=\frac{C_{\bar{\nu}}^{1/2}}{4-\bar{\nu}}\frac{1}{\Gamma\left(\frac{2-\bar{\nu}}{4-\bar{\nu}}\right)}\frac{2\sqrt{\mathcal{D}_{y}}}{\mathcal{D}_{x}}\left(C_{\bar{\nu}}^{1/2}\frac{4}{(4-\bar{\nu})^{2}}\frac{2\sqrt{\mathcal{D}_{y}}}{\mathcal{D}_{x}}\right)^{-\frac{1}{4-\bar{\nu}}}\frac{|x|^{\frac{1-\bar{\nu}}{2}}}{t^{\frac{3-\bar{\nu}}{2(4-\bar{\nu})}}}\nonumber\\&\times H_{1,2}^{2,0}\left[\left.\frac{4C_{\bar{\nu}}^{1/2}}{(4-\bar{\nu})^{2}}\frac{2\sqrt{\mathcal{D}_{y}}}{\mathcal{D}_{x}}\frac{|x|^{\frac{4-\bar{\nu}}{2}}}{t^{1/2}}\right|\left.\begin{array}{l} (\frac{5-\bar{\nu}}{2(4-\bar{\nu})},1/2)\\
(\frac{1}{4-\bar{\nu}},1),(-\frac{1}{4-\bar{\nu}},1)\end{array}\right.\right].
\end{eqnarray}

Solution (\ref{p1(x,t) final ex}) describes a subdiffusive behavior with the MSD 
\begin{eqnarray}
\left\langle x^{2}(t)\right\rangle=2\int_{0}^{\infty}dx\,x^{2}p_{1}(x,t)\simeq t^{\frac{2}{4-\bar{\nu}}},
\end{eqnarray}
where the transport exponent changes in the range $\frac{1}{2}<\frac{2}{4-\bar{\nu}}<\frac{2}{3}$. Note that the limiting case with $\bar{\nu}=0$ results in the continuous comb with the MSD $\left\langle x^{2}(t)\right\rangle\simeq t^{1/2}$.

\subsection{Special case with $\gamma(t)=\delta(t)$ and $\eta(t)=t^{-1/2}/\Gamma(1/2)$}

Next we consider the case with the kernels $\hat{\gamma}(s)=1$ and $\hat{\eta}(s)=s^{-1/2}$\footnote{For the continuous comb (\ref{diffusion-like eq delta memory Levy}), these memory functions give superdiffusion for the case $0<\nu<1$, and normal diffusion for $\nu=1$.}, which yields Eq.~(\ref{f(x,s) green}) in the form
\begin{eqnarray}\label{f(x,s) green ex1}
C_{\bar{\nu}}^{1/2}\frac{2\sqrt{\mathcal{D}_{y}}}{\mathcal{D}_{x}}s|x|^{-\bar{\nu}/2}\hat{G}(x,s)=\frac{\partial^{2}}{\partial x^{2}}\hat{G}(x,s).
\end{eqnarray}

Following the same procedure as above, we find the PDF $p_{1}(x,t)$ in the form
\begin{eqnarray}\label{p1(x,t) final ex1}
p_{1}(x,t)&=\frac{C_{\bar{\nu}}^{1/2}}{4-\bar{\nu}}\frac{1}{\Gamma\left(\frac{2-\bar{\nu}}{4-\bar{\nu}}\right)}\frac{2\sqrt{\mathcal{D}_{y}}}{\mathcal{D}_{x}}\left(C_{\bar{\nu}}^{1/2}\frac{4}{(4-\bar{\nu})^{2}}\frac{2\sqrt{\mathcal{D}_{y}}}{\mathcal{D}_{x}}\right)^{-\frac{1}{4-\bar{\nu}}}\frac{|x|^{\frac{1-\bar{\nu}}{2}}}{t^{\frac{3-\bar{\nu}}{4-\bar{\nu}}}}\nonumber\\&\times H_{1,2}^{2,0}\left[\left.C_{\bar{\nu}}^{1/2}\frac{4}{(4-\bar{\nu})^{2}}\frac{2\sqrt{\mathcal{D}_{y}}}{\mathcal{D}_{x}}\frac{|x|^{\frac{4-\bar{\nu}}{2}}}{t}\right|\left.\begin{array}{l} (\frac{1}{4-\bar{\nu}},1)\\
(\frac{1}{4-\bar{\nu}},1),(-\frac{1}{4-\bar{\nu}},1)\end{array}\right.\right]\nonumber\\&=\frac{C_{\bar{\nu}}^{1/2}}{4-\bar{\nu}}\frac{1}{\Gamma\left(\frac{2-\bar{\nu}}{4-\bar{\nu}}\right)}\frac{2\sqrt{\mathcal{D}_{y}}}{\mathcal{D}_{x}}\left(C_{\bar{\nu}}^{1/2}\frac{4}{(4-\bar{\nu})^{2}}\frac{2\sqrt{\mathcal{D}_{y}}}{\mathcal{D}_{x}}\right)^{-\frac{1}{4-\bar{\nu}}}\frac{|x|^{\frac{1-\bar{\nu}}{2}}}{t^{\frac{3-\bar{\nu}}{4-\bar{\nu}}}}\nonumber\\&\times H_{0,1}^{1,0}\left[\left.C_{\bar{\nu}}^{1/2}\frac{4}{(4-\bar{\nu})^{2}}\frac{2\sqrt{\mathcal{D}_{y}}}{\mathcal{D}_{x}}\frac{|x|^{\frac{4-\bar{\nu}}{2}}}{t}\right|\left.\begin{array}{l}\\
(-\frac{1}{4-\bar{\nu}},1)\end{array}\right.\right]\nonumber\\ 
&=\frac{C_{\bar{\nu}}^{1/2}}{4-\bar{\nu}}\frac{1}{\Gamma\left(\frac{2-\bar{\nu}}{4-\bar{\nu}}\right)}\frac{2\sqrt{\mathcal{D}_{y}}}{\mathcal{D}_{x}}\left(C_{\bar{\nu}}^{1/2}\frac{4}{(4-\bar{\nu})^{2}}\frac{2\sqrt{\mathcal{D}_{y}}}{\mathcal{D}_{x}}\right)^{-\frac{2}{4-\bar{\nu}}}\frac{|x|^{-\bar{\nu}/2}}{t^{\frac{2-\bar{\nu}}{4-\bar{\nu}}}}\nonumber\\&\times\exp\left(-C_{\bar{\nu}}^{1/2}\frac{4}{(4-\bar{\nu})^{2}}\frac{2\sqrt{\mathcal{D}_{y}}}{\mathcal{D}_{x}}\frac{|x|^{\frac{4-\bar{\nu}}{2}}}{t}\right),
\end{eqnarray}
which is normalized to one as well, and is of stretched exponential form. Here we used relations (\ref{H property2}) and (\ref{H relation exp}). The MSD now reads 
\begin{eqnarray}
\left\langle x^{2}(t)\right\rangle=2\int_{0}^{\infty}dx\,x^{2}p_{1}(x,t)\simeq t^{\frac{4}{4-\bar{\nu}}}.
\end{eqnarray}
This solution describes superdiffusion with the transport exponent ranging in the interval $1<\frac{4}{4-\bar{\nu}}<\frac{4}{3}$, which is enhanced diffusion in comparison to the solution in Eq.~(\ref{p1(x,t) final ex}). This is a Levy-like process, where the CTRW with spatio-temporal coupling takes place. The diffusion in the $x$ direction is enhanced due to the generalized compensation memory kernel $\eta(t)=\frac{t^{-1/2}}{\Gamma(1/2)}$\footnote{The presence of this compensation memory kernel in the continuous comb model (\ref{diffusion like eq on a comb0}) yields normal diffusion in the $x$ direction in comparison to the subdiffusive behavior with the transport exponent equals to $1/2$ in the classical comb model (\ref{classical comb}).}. The long jumps on the fractal comb are penalized by long waiting times. This mechanism leads to the stretched exponential behavior in the last line of Eq.~(\ref{p1(x,t) final ex1}), which eventually yields the finite MSD. The case with $\bar{\nu}=0$ recovers the result of the continuous comb with $\left\langle x^{2}(t)\right\rangle\simeq t$.

\section{Summary}

We considered L\'{e}vy processes in a generalized fractal comb model, which is derived from general properties of the Liouville equation, and we presented an exact analytical analysis of the solutions of equation (\ref{diffusion like eq on a comb0}) for the probability distribution function (PDF) for anomalous diffusion of particles for various realizations of the generalized comb model. Comb geometry is one of the most simple paradigms where anomalous diffusion can be realized in the framework of Markovian processes as in Eq.~(\ref{classical comb}). However, the intrinsic properties of the structure can destroy this Markovian transport. These effects violate the Markov consideration of Eq.~(\ref{classical comb}) and lead to the introduction of the memory $\eta(t)$, $\gamma(t)$, and spatial $\rho(x)$ kernels in Eq.~(\ref{diffusion like eq on a comb0}). The fractal structure of fingers, which is controlled by the spatial kernel $\rho(x)$ in the form of the power-law distributions in both real and Fourier spaces, leads to the L\'evy processes (L\'evy flights) and superdiffusion. In the former case, when the spatial kernel is defined in the real space, this effect is manifested by the Riesz fractional derivative of the order of $\alpha=(3+\nu)/2<2$, where $\nu$ is the fractal dimension of the fingers. This was observed for the first time in Ref.~\cite{iomin2}, where a qualitative analytical analysis has been suggested. In the present analysis, this problem is solved exactly and exact analytical solutions are obtained in the form of the Fox $H$-functions. In some extend, here we demonstrated an application of the Fox $H$-functions in solving anomalous diffusion equations. The interplay between the spatial kernel and the memory kernels, controlled by the heavy tail exponent $\mu$, is reflected in the transport exponent of the anomalous diffusion $\frac{2-\mu}{\alpha}$, such that when $2\mu+\nu<1$ there is superdiffusion. In the opposite case when $2\mu+\nu>1$ subdiffusion takes place. For the completeness of the analysis, cases with distributed order memory kernels are also investigated by employing the Tauberian theorem. As a result, we obtained ultra-slow diffusion. It is a robust slow process, which cannot be destroyed by the L\'evy flights. Finally, we considered the fractional structure of the fingers controlled by the Weierstrass function, which leads to the power-law  kernel in the Fourier space. A superdiffusive solution in Eq.~(\ref{p1(x,t) final ex1}) is found as well. It is expressed in the form of a stretched exponential function (\ref{p1(x,t) final ex1}). It is a special case, when the second moment exists for superdiffusion, since the L\'evy flights are interrupted by fingers-traps with the power-law waiting time PDF. In this case, the superdiffusive MSD is exactly calculated from the second moment $\langle x^2(t)\rangle=t^{\frac{4}{4-\nu}}$.

In conclusion, we discuss the question on the relation between fractal structures (like shown in Fig. 2) and fractional Riesz derivative as a reflection of the L\'evy dynamics. This problem has been considered in many studies \cite{blumen,iomin2,LM1,LM,nigmatulin,rutman,shlesinger}. Here, we also concern with a question what kind of information is neglected when random walk on quenched fractal structure is described by the Riezs fractional integral
\footnote{This relates to the link between fractal geometry and fractional integro-differentiation \cite{nigmatulin}, which is constituted in the procedure of averaging an extensive physical value
that is expressed by means of a smooth function over a Cantor set, which leads to fractional integration. However, as criticized in Ref. \cite{rutman}, the Cantor set ``as a memory function allows for no convolution''. In its eventual form, the link has been presented in Ref. \cite{nigmatulin} as an averaging procedure over the log periodicity of the fractal.}. The answer is as follows. The fractal structure, like in Fig. 2 can be described for example by the Weierstrass function, which depends on two parameters $l$ and $b$, which lead to the scaling in Eq.~(\ref{Weierstrass3}) and to the log periodicity, and as well as to the fractal volume with the fractal dimension $\bar{\nu}=\log{b}/\log{l}$. However, the asymptotic approximation contains only the fractal volume, while the self-similarity and log periodicity properties are already lost. This expression is explicitly obtained in \ref{WF}. In this case a regular fractal is considered as a random fractal with the fractal volume $|x|^\nu$. It should be admitted that in Sec. 2, our construction of the Riesz space fractional integration by means of the power law kernel $\rho(x)$ is exact. In this sense, our analytical description of the L\'evy process is exact, however, its relation to the Cantor set of the fingers is just illustrative. A rigorous coarse-grained procedure, which  relates the fractal structure of the comb fingers to the Riesz fractional derivative has been established in Ref. \cite{iomin2}. The situation changes dramatically in Sec. 3, where the Weierstrass function describes rigorously the fractal comb. However, in our analytical treatment we use only its asymptotic approximation \cite{blumen} to obtain fractional integro-differentiation. As admitted above, in this case all information on self-similarity and log periodicity is lost.

\ack

TS acknowledges the hospitality and support from the Max-Planck Institute for the Physics of Complex Systems in Dresden, Germany. AI was supported by the Israel Science Foundation (ISF-1028). VM is supported by Grants No. FIS 2012-32334 by the Ministerio de Economia y Competitividad and by SGR 2013-00923 by the Generalitat de Catalunya.

\appendix

\section{Solution of Eqs.~(\ref{diffusion-like eq delta memory Levy}) and (\ref{diffusion-like eq two power law memory Levy})}\label{app solution}

We note, first, that Eqs.~(\ref{diffusion-like eq delta memory Levy}) is a particular case of Eq.~(\ref{diffusion-like eq two power law memory Levy}), which is a general form of a space-time fractional diffusion equation
\begin{eqnarray}\label{Eq.100}
{_{C}}D_{t}^{\lambda}
p_{1}(x,t)=\mathcal{D}_{\lambda,\alpha}\frac{\partial^\alpha}{\partial
|x|^\alpha}p_{1}(x,t), \quad t>0, \quad -\infty<x<+\infty,
\end{eqnarray}
where ${_{C}}D_{t}^{\lambda}$ is the Caputo time fractional derivative (\ref{Caputo_derivative}) of order $0<\lambda<1$, $\frac{\partial^\alpha}{\partial
|x|^\alpha}$ is the Riesz space fractional derivative of order $1<\alpha<2$, and $\mathcal{D}_{\lambda,\alpha}$ is the generalized diffusion coefficient with physical dimension $\left[\mathcal{D}_{\lambda,\alpha}\right]=\mathrm{m}^{\alpha}\mathrm{s}^{-\lambda}$. The boundary conditions at infinities are
\begin{eqnarray}\label{Eq.200}
p_{1}(\pm\infty,t)=0, \quad \frac{\partial}{\partial x}p_{1}(\pm\infty,t)=0, \quad t>0,
\end{eqnarray}
while the initial condition is
\begin{eqnarray}\label{Eq.300}
p_{1}(x,0)=\delta(x), \quad -\infty<x<+\infty.
\end{eqnarray}
Applying the Fourier-Laplace transform in Eq.~(\ref{Eq.100}), and accounting the initial condition (\ref{Eq.300}) and the boundary conditions (\ref{Eq.200}), one finds
\begin{eqnarray}\label{laplacefourier100}
\tilde{\hat{p}}_{1}(\kappa,s)=\frac{s^{\lambda-1}}{s^{\lambda}+\mathcal{D}_{\lambda,\alpha}|\kappa|^\alpha
}.
\end{eqnarray}
Here we use the property of the Laplace transform for the Caputo derivative \cite{samko}
\begin{eqnarray}\label{laplace Caputo}
\mathcal{L}\left[{_{C}}D_{t}^{\lambda}f(t)\right]=s^{\lambda}\mathcal{L}\left[f(t)\right]-s^{\lambda-1}f(0).
\end{eqnarray}
From the inverse Laplace transform, by employing formula \cite{mainardi_book}
\begin{eqnarray}
\label{ML three Laplace}
\mathcal{L}\left[t^{\beta-1}E_{\alpha,\beta}(\pm at^{\alpha})\right]=
\frac{s^{\alpha-\beta}}{s^{\alpha}\mp a},
\end{eqnarray}
for $\Re(s)>|a|^{1/\alpha}$, where $E_{\alpha,\beta}(z)$ is the two parameter Mittag-Leffler function (\ref{ML three}), it follows
\begin{eqnarray}\label{inverselaplace100}
\tilde{p}_{1}(\kappa,t)=E_{\lambda}\left(-\mathcal{D}_{\lambda,\alpha}t^{\lambda}|\kappa|^\alpha
\right).
\end{eqnarray}
Here $E_{\lambda}(z)$ is the one parameter Mittag-Leffler function (\ref{ML three}). From relations (\ref{HML}) and (\ref{H property1}), and the Fourier transform formula (\ref{cosine H}), one obtains the solution of Eq.~(\ref{Eq.100}) in terms of the Fox $H$-function (\ref{H_integral}) \cite{MPS,physica a 2012}:
\begin{eqnarray}\label{special case}
p_{1}(x,t)&=\frac{2}{2\pi}\int_{0}^{\infty}d\kappa\,\cos(\kappa x)H_{1,2}^{1,1}\left[\mathcal{D}_{\lambda,\alpha}t^{\lambda}|\kappa|^\alpha
\left|\begin{array}{l}
    (0,1)\\
    (0,1),(0,\lambda)
  \end{array}\right.\right]\nonumber\\&=\frac{1}{\alpha\pi}\int_{0}^{\infty}d\kappa\,\cos(\kappa x)H_{1,2}^{1,1}\left[\left(\mathcal{D}_{\lambda,\alpha}t^{\lambda}\right)^{1/\alpha}|\kappa|
  \left|\begin{array}{l}
      (0,1/\alpha)\\
      (0,1/\alpha),(0,\lambda/\alpha)
    \end{array}\right.\right]\nonumber\\&=\frac{1}{\alpha
|x|}H_{3,3}^{2,1}\left[\frac{|x|}{\left(\mathcal{D}_{\lambda,\alpha}t^{\lambda}\right)^{1/\alpha}}
\left|\begin{array}{l}
    (1,\frac{1}{\alpha}),(1,\frac{\lambda}{\alpha}),(1,\frac{1}{2})\\
    (1,1),(1,\frac{1}{\alpha}),(1,\frac{1}{2})
  \end{array}\right.\right].
\end{eqnarray}
From the solution (\ref{special case}), by using relation (\ref{integral of H}), we obtain the fractional moments (\ref{fractional moments}) \cite{physica a 2012}
\begin{eqnarray}\label{moments case}
&\left\langle |x|^{q}(t)\right\rangle=\frac{2}{\alpha}\int_{0}^{\infty}dx\,x^{q-1}H_{3,3}^{2,1}\left[\frac{x}{\left(\mathcal{D}_{\lambda,\alpha}t^{\lambda}\right)^{1/\alpha}}
\left|\begin{array}{l}
    (1,\frac{1}{\alpha}),(1,\frac{\lambda}{\alpha}),(1,\frac{1}{2})\\
    (1,1),(1,\frac{1}{\alpha}),(1,\frac{1}{2})
  \end{array}\right.\right]\nonumber\\ &=\frac{2}{\alpha}\left(\mathcal{D}_{\lambda,\alpha}t^{\lambda}\right)^{q/\alpha}\theta(-q)
=\frac{4}{\alpha}\cdot\frac{\Gamma\left(q\right)\Gamma(1+q/\alpha)\Gamma(-q/\alpha)}{\Gamma(q/2)\Gamma(-q/2)}\cdot
\frac{\left(\mathcal{D}_{\lambda,\alpha}t^\lambda\right)^{q/\alpha}}{\Gamma\left(1+\frac{\lambda q}{\alpha}\right)},\nonumber\\
\end{eqnarray}
where we apply $\Gamma(1-z)\Gamma(z)=\frac{\pi}{\sin\left(\pi z\right)}$ \cite{erdelyi}, and where, for the current example,
\begin{eqnarray}\label{theta ex}
\theta(q)=\frac{\Gamma(1+q)\Gamma(1+q/\alpha)\Gamma(-q/\alpha)}{\Gamma(-q/2)\Gamma(1+\lambda q/\alpha)\Gamma(1+q/2)} =\frac{2\Gamma(q)\Gamma(1+q/\alpha)\Gamma(-q/\alpha)}{\Gamma(-q/2)\Gamma(1+\lambda q/\alpha)\Gamma(q/2)}.\nonumber\\
\end{eqnarray}

\section{Fox $H$-function and Mittag-Leffler functions}

\subsection{Fox $H$-function}\label{app_fox}

A detailed description of the Fox H-function and its application can be found in Refs.~\cite{saxena book,MH2008}.

The Fox $H$-function is defined in terms of the Mellin-Barnes integral
\begin{eqnarray}
H_{p,q}^{m,n}\left[z\left|\begin{array}{l}(a_1,A_1),\ldots,(a_p,A_p)\\
(b_1,B_1),\ldots,b_q,B_q)\end{array}\right.\right]=\frac{1}{2\pi\imath}\int_{\Omega}ds\,\theta(s)z^{-s},
\label{H_integral}
\end{eqnarray}
where
\begin{eqnarray}
\theta(s)=\frac{\prod_{j=1}^{m}\Gamma(b_j+B_js)\prod_{j=1}^{n}\Gamma(1-a_j-A_js)}{
\prod_{j=m+1}^{q}\Gamma(1-b_j-B_js)\prod_{j=n+1}^{p}\Gamma(a_j+A_js)},
\end{eqnarray}
with $0\leq n\leq p$, $1\leq m\leq q$, $a_i,b_j \in C$, $A_i,B_j\in R^{+}$, $i=1,
\ldots,p$, and $j=1,\ldots,q$. The contour $\Omega$, starting at $c-i\infty$ and
ending at $c+i\infty$, separates the poles of the function $\Gamma(b_j+B_js)$, $j
=1,\ldots,m$ from those of the function $\Gamma(1-a_i-A_is)$, $i=1,\ldots,n$. 

The Fox $H$-function is symmetric in the pairs $(a_1, A_1),\ldots,(a_n, A_n)$, likewise
$(a_{n+1}, A_{n+1}),\ldots,(a_p, A_p)$; in $(b_1, B_1),\ldots,(b_m, B_m)$ and $(b_{
m+1},B_{m+1}),\ldots,(B_q,B_q)$. 

The Fox $H$-function has the following properties
\begin{eqnarray}
\label{H property1}
H_{p,q}^{m,n}\left[z^{\delta}\left|\begin{array}{l}(a_p,A_p)\\(b_q,B_q)\end{array}
\right.\right]=\frac{1}{\delta}H_{p,q}^{m,n}\left[z\left|\begin{array}{l}(a_p,A_p/
\delta)\\(b_q,B_q/\delta)\end{array}\right.\right],
\end{eqnarray}
where $\delta>0$,
\begin{eqnarray}
&\label{H property2}
H_{p,q}^{m,n}\left[z\left|\begin{array}{l}(a_1,A_1),\dots,(a_{p-1},A_{p-1}),(b_{1},B_{1})\\(b_{1},B_{1}),(b_{2},B_{2}),\dots,(b_q,B_q)\end{array}
\right.\right]\nonumber\\&=H_{p-1,q-1}^{m-1,n}\left[z\left|\begin{array}{l}(a_1,A_1),\dots,(a_{p-1},A_{p-1})\\(b_{2},B_{2}),\dots,(b_q,B_q)\end{array}
\right.\right],
\end{eqnarray}
where $m\ge1$, and $p>n$.

The Mellin transform of the Fox $H$-function is given by
\begin{eqnarray}
\int_0^{\infty}dx\,x^{\xi-1}H_{p,q}^{m,n}\left[ax\left|\begin{array}{l}(a_p,A_p)\\
(b_q,B_q)\end{array}\right.\right]=a^{-\xi}\theta(\xi),
\label{integral of H}
\end{eqnarray}
where $\theta(\xi)$ is defined in relation (\ref{H_integral}). 

The Mellin-cosine transform of the Fox $H$-function is given by
\begin{eqnarray}
\label{cosine H}
&\int_{0}^{\infty}{d}\kappa\,\kappa^{\rho-1}\cos(\kappa x)H_{p,q}^{m,n}\left[a\kappa^{\delta}\left|
\begin{array}{l}(a_p,A_p)\\(b_q,B_q)\end{array}\right.\right]
\nonumber\\&=\frac{\pi}{x^\rho}H_{q+1,p+2}^{n+1,m}\left[\frac{x^\delta}{a}\left|
\begin{array}{l}(1-b_q,B_q),(\frac{1+\rho}{2},\frac{\delta}{2})\\(\rho,\delta),
(1-a_p,A_p),(\frac{1+\rho}{2},\frac{\delta}{2})\end{array}\right.\right],
\end{eqnarray}
where
\begin{eqnarray}
&\Re\left(\rho+\delta \min_{1\leq j\leq m}\left(\frac{b_j}{B_j}\right)\right)>1,
\quad x^\delta>0,\nonumber\\
&\Re\left(\rho+\delta \max_{1\leq j\leq n}\left(\frac{a_j-1}{A_j}\right)\right)<
\frac{3}{2}, \quad |\arg(a)|<\pi\alpha/2,\nonumber\\
&\alpha=\sum_{j=1}^{n}A_j-\sum_{j=n+1}^{p}A_j+\sum_{j=1}^{m}B_j-\sum_{j=m+1}^{q}
B_j>0.\nonumber
\end{eqnarray}

The following Laplace transform formula is true for the Fox $H$-function
\begin{eqnarray}
\mathcal{L}^{-1}\left[s^{-\rho}H_{p,q}^{m,n}\left[as^{\sigma}\left|\begin{array}{l}(a_p,A_p)\\
(b_q,B_q)\end{array}\right.\right]\right]=t^{\rho-1}H_{p+1,q}^{m,n}\left[\frac{a}{t^{\sigma}}\left|\begin{array}{l}(a_p,A_p),(\rho,\sigma)\\
(b_q,B_q)\end{array}\right.\right].\nonumber\\
\label{H_laplace}
\end{eqnarray}

The Bessel function of third kind $K_{\nu}(z)$ is a special case of the Fox $H$-function
\begin{eqnarray}
H_{0,2}^{2,0}\left[\frac{z^{2}}{4}\left|\begin{array}{l} \\
(\frac{a+\nu}{2},1),(\frac{a-\nu}{2},1)\end{array}\right.\right]=2\left(\frac{z}{2}\right)^{a}K_{\nu}(z).
\label{HK relation}
\end{eqnarray}

Series representation of modified Bessel function of the second kind is given by
\begin{eqnarray}
K_{\nu}(z)&\simeq\frac{\Gamma(\nu)}{2}\left(\frac{z}{2}\right)^{-\nu}\left[1+\frac{z^{2}}{4(1-\nu)}+\dots\right]\nonumber\\&+\frac{\Gamma(-\nu)}{2}\left(\frac{z}{2}\right)^{\nu}\left[1+\frac{z^{2}}{4(\nu+1)}+\dots\right], \quad z\rightarrow0, \quad \nu\notin Z.
\label{K series}
\end{eqnarray}

For special case of parameters of the Fox $H$-function, one obtains
\begin{eqnarray}
H_{0,1}^{1,0}\left[z\left|\begin{array}{l} \\
(b,B)\end{array}\right.\right]=B^{-1}z^{b/B}\exp\left(-z^{1/B}\right).
\label{H relation exp}
\end{eqnarray}

\subsection{Mittag-Leffler functions}

The two parameter Mittag-Leffler function is defined by \cite{mainardi_book}
\begin{equation}\label{ML three}
E_{\alpha,\beta}(z)=\sum_{k=0}^{\infty}\frac{z^{k}}{\Gamma(\alpha
k+\beta)}.
\end{equation}
The one parameter Mittag-Leffler function $E_{\alpha}(z)$ is a special case of the two parameter Mittag-Leffler function if we set $\beta=1$. 

The two parameter Mittag-Leffler function (\ref{ML three}) is a special case of the
Fox $H$-function \cite{saxena book}
\begin{equation}
\label{HML}
E_{\alpha,\beta}(-z)=H_{1,2}^{1,1}\left[z\left|
\begin{array}{l}(0,1)\\(0,1),(1-\beta,\alpha)\end{array}\right.\right].
\end{equation}

\section{Weierstrass function}\label{WF}

Here we will show that the discrete, fractal distribution of fingers, can be constructed by means of the 
Weierstrass function. We will follow the approach recently used in \cite{sandev iomin kantz}, where it is shown that the fractal structure of backbones corresponds to the Weierstrass function inside the backbones. Let us consider Eq.~(\ref{diffusion like eq on a comb0}), where the last term is given by $\mathcal{D}_{y}\frac{\partial^{2}}{\partial
y^{2}}\sum_{j=1}^{\infty}w_{j}\delta(x-l_{j})P(x,y,t)$, i.e., we investigate the following equation
\begin{eqnarray}\label{diffusion like eq on a comb weierstrass}
\int_{0}^{t}dt'\,\gamma(t-t')\frac{\partial}{\partial t'}P(x,y,t')
&=\mathcal{D}_{x}\delta(y)\int_{0}^{t}dt'\,\eta(t-t')\frac{\partial^{2}}{\partial x^{2}}P(x,y,t')\nonumber\\&+\mathcal{D}_{y}\sum_{j=1}^{\infty}w_{j}\delta(x-l_{j})\frac{\partial^{2}}{\partial
y^{2}}P(x,y,t).
\end{eqnarray}
The last term in this equation means that the diffusion along the $y$ axis occurs on
infinite number of fingers located at $x=l_{j}$, $j=1,2,...$, $0\leq
l_{1}<l_{2}<\dots<l_{N}<\dots$, at positions $x$ which belong to the fractal set $S_{\nu}$ with fractal dimension $0<\nu<1$, and $w_{j}$ are structural constants such that 
\begin{eqnarray}\label{sum w}
\sum_{j=1}^{\infty}w_{j}=1.
\end{eqnarray}
The summation in the last term of Eq.~(\ref{diffusion like eq on a comb weierstrass}), is a summation over a fractal set $S_{\nu}$.
In order to obtain the Weierstrass function we follow the procedure given in \cite{sandev iomin kantz,shlesinger}. Therefore, we use that
$w_{j}=\frac{l-b}{b}\left(\frac{b}{l}\right)^{j}$, where $l,b>1$, $l-b\ll b$ ($l$ and $b$ are dimensionless scale parameters), from where we find
\begin{equation}\label{w_j for Weierstrass}
\sum_{j=1}^{\infty}w_{j}=\frac{l-b}{l}\sum_{j=0}^{\infty}\left(\frac{b}{l}\right)^{j}=1,
\end{equation}
as it should be for the structural constants (\ref{sum w}). From (\ref{eq in FF space}) and (\ref{w_j for Weierstrass}) it follows
\begin{equation}\label{Weierstrass}
\Psi(z)=\frac{l-b}{b}\sum_{j=1}^{\infty}\left(\frac{b}{l}\right)^{j}\exp\left(i\frac{z}{l^{j}}\right),
\end{equation}
where $l_{j}=L/l^{j}$, $z=\left(\kappa_{x}-\kappa_{x}{'}\right)L$, and $l_{1}=L=1$. From here one obtains \cite{sandev iomin kantz}
\begin{equation}\label{Weierstrass2}
\Psi(z/l)=\frac{l}{b}\Psi(z)-\frac{l-b}{b}\exp\left(i\frac{z}{l}\right),
\end{equation}
and by neglecting the last term ($l-b\ll b$), the following scaling is found
\begin{equation}\label{Weierstrass3}
\Psi(z/l)\simeq\frac{l}{b}\Psi(z).
\end{equation}
This means that $\Psi(z)\sim\frac{1}{z^{1-\bar{\nu}}}$, where
$\bar{\nu}=\log{b}/\log{l}$, $0<\bar{\nu}<1$, is the fractal dimension. From here, for the last term in (\ref{diffusion like eq on a comb weierstrass}) we have
\begin{eqnarray}\label{eq in FF space2}
\mathcal{D}_{y}\kappa_{y}^{2}\frac{1}{2\pi}\int_{-\infty}^{\infty}d\kappa_{x}{'}\frac{\tilde{P}(\kappa_{x}{'},\kappa_{y},t)}{|\kappa_{x}-\kappa_{x}{'}|^{1-\bar{\nu}}},
\end{eqnarray}
which is the Riesz fractional integral \cite{samko} in the reciprocal Fourier space.


\section*{References}

\begin{thebibliography} {}

\bibitem{arkhincheev1}
Arkhincheev V E 2007 {\it Chaos} {\bf17} 043102 

\bibitem{arkhincheev}
Arkhincheev  V E and Baskin E M 1991 {\it Sov. Phys. JETP} {\bf73} 161

\bibitem{nature06948} 
Barthelemy P, Bertolotti J and Wiersma D S 2008 {\it Nature} {\bf453} 495

\bibitem{baskin iomin prl}
Baskin E and Iomin A 2004 {\it Phys. Rev. Lett.} {\bf93} 120603

\bibitem{berry}
Berry M V and Lewis Z V 1980 {\it Proc. R. Soc. London Ser. A} {\bf370} 459

\bibitem{blumen}
Blumen A,  Klafter J, and  Zumofen G 1985 in {\it Fractals in Physics}, edited by L. Pietronero and E. Tosatti  (Amsterdam: North-Holland), p. 399.

\bibitem{report}
Bouchaud J -P and Georges A 1990 {\it Phys. Rep.} {\bf195} 127

\bibitem{BurPRE81} 
Burioni R, Caniparoli L and Vezzani A 2010 {\it Phys. Rev. E} {\bf81} 060101R

\bibitem{BurPRE89} 
Burioni R, Ubaldi E and Vezzani A 2014 {\it Phys. Rev. E} {\bf89} 022135

\bibitem{cassi}
Cassi D and Regina S 1996 {\it Phys. Rev. Lett.} {\bf76} 2914\\
Baldi G, Burioni R and Cassi D 2004 {\it Phys. Rev. E} {\bf70} 031111

\bibitem{chechkin2}
Chechkin A V, Gorenflo R and Sokolov I M 2002 {\it Phys. Rev. E} {\bf66} 046129

\bibitem{chechkin}
Chechkin A V, Klafter J and Sokolov I M 2003 {\it Europhys. Lett.} {\bf63} 326

\bibitem{lenzi}
da Silva L R, Tateishi A A, Lenzi M K, Lenzi E K and da Silva P C2009 {\it Brazilian J. Phys.} {\bf39} 438

\bibitem{erdelyi} 
Erdelyi A, Magnus W, Oberhettinger F and Tricomi F G 1955 {\it Higher Transcedential Functions} vol 3 (New York: McGraw-Hill)

\bibitem{falconer}
Falconer K 1990 {\it Fractal Geometry} (New York: John Wiley \& Sons Ltd)

\bibitem{fedotov}
Fedotov S and Mendez V 2008 {\it Phys. Rev. Lett.} {\bf101} 218102

\bibitem{feller}
Feller W 1968 {\it An Introduction to Probability Theory and Its Applications} vol II (New York: John Wiley \& Sons Ltd)

\bibitem{book integrals}
Gradshteyn I S and Ryzhik I M 2007 {\it Table of Integrals, Series, and Products} (San Diego: Academic Press)

\bibitem{iomin2}
Iomin A 2011 {\it Phys. Rev. E} {\bf83} 052106

\bibitem{iomin3}
Iomin A 2012 {\it Phys. Rev. E} {\bf86} 032101

\bibitem{iomin baskin} 
Iomin A and Baskin E 2005 {\it Phys. Rev. E} {\bf71} 061101

\bibitem{iomin}
Iomin A and Mendez V 2013 {\it Phys. Rev. E} {\bf88} 012706

\bibitem{IS2012} 
Iomin A and Sokolov I M 2012 {\it Phys. Rev. E} {\bf86} 022101

\bibitem{kochubei}
Kochubei A N 2011 {\it Integr. Equ. Oper. Theory} {\bf71} 583

\bibitem{LM1}
Le M\'ehaute A 1990 \textit{Les g\'eometries fractales} (Paris: Hermes)

\bibitem{LM}
Le M\'ehaute A, Nigmatullin R R and Nivanen L 1998 {\it Fleches du Temps et Geometric Fractale} (Paris: Hermes).

\bibitem{mainardi_book}
Mainardi F 2010 {\it Fractional Calculus and Waves in Linear Viscoelesticity: An introduction to Mathematical Models} (London: Imperial College Press)

\bibitem{MPS}
Mainardi F, Pagnini G and Saxena R K 2005 {\it J. Comput. Appl. Math.} {\bf178} 321

\bibitem{havlin}
Matan O, Havlin S and Staufler D 1989 {\it J. Phys. A: Math. Gen.} {\bf22} 2867

\bibitem{MH2008} 
Mathai A M and Haubold H J 2008 {\it Special Functions for Applied Scientists} (New York: Springer)

\bibitem{saxena book}
Mathai A M, Saxena R K and Haubold H J 2010 {\it The $H$-function: Theory and Applications} (New York: Springer)

\bibitem{mendez}
Mendez V and Iomin A 2013 {\it Chaos Solitons \& Fractals} {\bf53} 46

\bibitem{mich2015}  
Mendez V, Iomin A, Campos D and Horsthemke W 2015 {\it Phys. Rev. E} {\bf92} 062112 

\bibitem{metzler cppc}
Metzler R, Jeon J- H, Cherstvy A G and Barkai E 2014 {\it Phys. Chem. Chem. Phys.} {\bf16} 24128
 
\bibitem{metzler report}
Metzler R and Klafter J 2000 {\it Phys. Rep.} {\bf339} 1\\ Metzler R and Klafter J 2004 {\it J. Phys. A: Math. Gen.} {\bf37} R161

\bibitem{nigmatulin}
Nigmatulin R R 1992 {\it Theor. Math. Phys.} \textbf{90} 245


\bibitem{rutman}
Rutman R S 1994 {\it Teoret. Mat. Fiz.} {\bf100} 476

\bibitem{samko}
Samko S G, Kilbas A A and Marichev O I 1993 {\it Fractional Integrals and Derivatives: Theory and Applications} (Philadelphia: Gordon and Breach Science Publishers)

\bibitem{sandev draft}
Sandev T, Chechkin A, Kantz H and Metzler R 2015 {\it Fract. Calc. Appl. Anal.} {\bf18} 1006

\bibitem{sandev iomin kantz}
Sandev T, Iomin A and Kantz H 2015 {\it Phys. Rev. E} {\bf91} 032108

\bibitem{joint MMNP}

Sandev T, Iomin A, Kantz H, Metzler R and Chechkin A 2016 {\it Math. Model. Nat. Phenom.} {\bf11} 18

\bibitem{sandev jpa2011}
Sandev T, Metzler R and Tomovski Z 2011 {\it J. Phys. A: Math. Theor.} {\bf44} 255203

\bibitem{baklanov}
Shamiryan D, Baklanov M R, Lyons P, Beckx S, Boullart W and Maex K 2007 {\it Colloids and Surfaces A: Physicochem. Eng. Aspects} {\bf300} 111

\bibitem{shlesinger}
Shlesinger M F 1974 {\it J. Stat. Phys.} {\bf10} 421

\bibitem{Sinai}  
Sinai Ya G 1982 {\it Theory Probab. Appl.} {\bf27} 256

\bibitem{physica a 2012}
Tomovski Z, Sandev T, Metzler R and Dubbeldam J 2012 {\it Physica A} {\bf391} 2527

\bibitem{weiss}
Weiss G H and Havlin S 1986 {\it Physica A} {\bf134} 474
 
\bibitem{west}
West B J, Grigolini P, Metzler R and Nonnenmacher T F 1997 {\it Phys. Rev. E} {\bf55} 99\\ Jespersen S, Metzler R and Fogedby H C 1999 {\it Phys. Rev. E} {\bf59} 2736

\bibitem{white}
White S R and Barma M 1984 {\it J. Phys. A: Math. Gen.} {\bf17} 2995

\end{thebibliography}
\end{document}